\documentclass[aps,pra,twocolumn,groupedaddress,amssymb,amsfonts,showpacs]{revtex4}

\input{Qcircuit}

\newcommand{\<}{\langle}
\renewcommand{\>}{\rangle}

\newcommand{\C}{\mathbb{C}}

\newcommand{\cA}{\mathcal{A}}
\newcommand{\cB}{\mathcal{B}}
\newcommand{\cH}{\mathcal{H}}
\newcommand{\cK}{\mathcal{K}}

\newcommand{\s}{\mathrm{span}}

\newcommand{\R}{\mathbb{R}}

\newtheorem{definition}{Definition}
\newtheorem{theorem}{Theorem}
\newtheorem{lemma}{Lemma}
\newtheorem{cor}{Corollary}

\begin{document}

\title{Speed-up via Quantum Sampling}

\author{Pawel Wocjan}
\email{wocjan@eecs.ucf.edu}
\affiliation{School of Electrical Engineering and Computer Science,
University of Central Florida, Orlando, FL~32816, USA.}

\author{Anura Abeyesinghe}
\affiliation{School of Electrical Engineering and Computer Science,
University of Central Florida, Orlando, FL~32816, USA.}

\begin{abstract}The Markov Chain Monte Carlo method is at the heart of efficient approximation schemes for a wide range of problems in combinatorial enumeration and statistical physics.  It is therefore very natural and important to determine whether quantum computers can speed-up classical mixing processes based on Markov chains.  To this end, we present a new quantum algorithm, making it possible to prepare a quantum sample, i.e., a coherent version of the stationary distribution of a reversible Markov chain. Our algorithm has a significantly better running time than that of a previous algorithm based on adiabatic state generation. We also show that our methods provide a speed-up over a recently proposed method for obtaining ground states of (classical) Hamiltonians.
\end{abstract}

\date{September 9, 2008}


\maketitle

\section{Introduction}
Randomization plays a crucial role in the design of efficient algorithms for computing approximate solutions to problems in combinatorial enumeration and statistical physics that are known to be \#P-complete.  Important examples are randomized polynomial-time approximation schemes for evaluating the permanent of a non-negative matrix \cite{Vigoda}, the volume of a convex polytope \cite{Dyer}, and the partition functions of the monomer-dimer and ferromagnetic Ising systems \cite{Jerrum1,Jerrum2}.  The centerpiece of all these algorithms is the Markov Chain Monte Carlo (MCMC) method, making it possible to approximately sample from a particular probability distribution $\pi$ over a large set $\Omega$.  

In the MCMC method, one constructs a sparse, ergodic Markov chain (stochastic matrix) $P$ on the state space $\Omega$ such that its stationary distribution $\pi$ is the desired probability distribution and then, starting from some initial state $x$, repeatedly applies $P$ so that the resulting probability distribution over $\Omega$ after $\tau$ steps is sufficiently close to $\pi$. The required number of step $\tau$ is referred to as mixing time.  Bounding the mixing time of the Markov chain is often the major technical hurdle in proving the running time of the overall algorithm.  This problem can be reduced to estimating the spectral gap $\delta$ of $P$ as described in the following.

We refer the reader to \cite{Sinclair,Jerrum} for more details on Markov chains. Let $P$ be an ergodic (i.e., irreducible and aperiodic) reversible Markov chain with finite state space $\Omega$ and stationary distribution $\pi=\big(\pi(x)\big)_{x\in\Omega}$. Let $P^{(t)}(x,y)$ denote the $t$-step transition probability from $x$ to $y$ and $P^{(t)}(x,\cdot)$ the probability distribution if we start in $x$ and apply the Markov chain $t$ times. To determine how fast the stationary distribution is approached from some initial probability distribution, we have to look at the spectral properties of the transition matrix $P$.

It follows by the Perron-Frobenius theorem that stationary distribution $\pi$ is the unique (left) eigenvector of $P$, i.e., $\pi^T P = \pi^T$, with associated eigenvalue $\lambda_0=1$.  Let $\{\lambda_j \, : \, 1\le j \le N-1\}$, $\lambda_j\in\R$, denote the remaining eigenvalues (not necessary distinct), where $N=|\Omega|$.  It also follows that these eigenvalues satisfy $|\lambda_j| < 1$ for $1\le j\le N-1$.  Let us order the eigenvalues such that
\begin{equation*}
1=\lambda_0 > |\lambda_1| \ge |\lambda_2| \ge \ldots \ge |\lambda_{N-1}|\ge 0\,.
\end{equation*}
Let $\delta$ denote the spectral gap of $P$, i.e., $\delta = 1-|\lambda_1|$.  The rate of convergence to the stationary distribution is governed by the spectral gap \cite[page 61]{Sinclair}.

The variation distance from initial state $x$ is 
\begin{equation*}
d^{(t)}(x) = \frac{1}{2} \sum_{y\in\Omega} |P^{(t)}(x,y)-\pi(y)|\,.
\end{equation*}
For $\epsilon\in (0,1)$, let
\begin{eqnarray*}
\tau_\epsilon(x) & = & \min\{ t \, : d^{(t')}(x) \le \epsilon \mbox{ for all $t'\ge t$}\} \\
\tau_\epsilon    & = & \max_{x\in\Omega} \tau_{\epsilon}(x)\,.
\end{eqnarray*}
These quantities satisfy the following inequalities:
\begin{eqnarray*}
\frac{1}{2 \delta} \log (2\epsilon)^{-1} & \le & \tau_{\epsilon} \\
\tau_{\epsilon}(x)                       & \le & \frac{1}{\delta} \Big( \log \big(\pi(x)\big)^{-1} + \log \epsilon^{-1} \Big) 
\end{eqnarray*}

Given the fact the MCMC method is at the heart of so many efficient classical algorithms, it is very natural and important to determine to what extent quantum computers can speed-up classical mixing. It possible to prepare the quantum sample $|\pi\>$, i.e., a coherent version of the stationary distribution where the amplitudes are given by the square roots of the probabilities of the states by quantized Markov chains \cite{Szegedy,Magniez}.  The running time of this algorithm is $O(1/\sqrt{\delta\, \pi(x)})$.  Basically, this algorithm is Grover search where marking of the target state is the reflection around the quantum sample $|\pi\>$.  This reflection can be realized with cost $O(1/\sqrt{\delta})$.

Unfortunately, this running time is too high for applications where the state space $\Omega$ is exponentially large and $\pi(x)$ can be exponentially small.  The question whether a quantum speed-up to $O(1/\sqrt{\delta}\, \log(1/\pi(x))$ is possible has been examined in \cite{Richter1}. The author proposes a method based on quantum walks that decohere under repeated randomized measurements to attack this problem.   He shows that this speed-up is indeed achievable for the decoherent quantum walk on a periodic lattice $\mathbb{Z}_n^d$.  However, the question whether this speed-up is achievable for arbitrary Markov chains remains an important open problem.

In this paper, we propose a different method for speeding-up classical mixing processes.  We show how to efficiently prepare the quantum sample $|\pi\>$ provided that we have a sequence of slowly-varying Markov chains in the following sense: (i) there are Markov chains $P_0,P_1,\ldots,P_r=P$ with stationary distributions $\pi_0,\pi,\ldots,\pi_r$ such that distributions of adjacent Markov chains are sufficiently close and (ii) the quantum sample $|\pi_0\>$ can be prepared efficiently.  The idea of quantum state generation based slowly-varying Markov chains was proposed in \cite{TaShma}, where adiabatic techniques were used to create the quantum samples $|\pi_i\>$ sequentially.  The resulting running time is $\Omega(1/\delta)$, i.e., it does not provided the speed-up that would reduce $\delta$ to $\sqrt{\delta}$. 

We improve the running time by (a) using Szegedy's quantum walk operators instead of Hamiltonians derived from the Markov chains and (b) preparing the intermediate quantum samples by amplitude amplification.  The resulting running time is worse than Richter's conjectured running time $O(1/\sqrt{\delta}\, \log(1/\pi(x))$.  However, in some situations, our running time is better than the classical lower bound on the mixing time. 

We then apply our method to the special case of simulated annealing.  We obtain a better running time than that of a recently proposed method for this purpose \cite{Somma,Somma2}. 

Their method makes use of Szegedy's quantum walk operators and the quantum Zeno effect.  It yields a quadratic speed-up with respect to the spectral gap and provides an overall speed-up with respect to the classical algorithm for simulated annealing. However, if one applied their method to the general case considered in \cite{TaShma}, one would also obtain the quadratic speed-up with respect to the spectral gap but not the {\em overall} speed-up.  The problem is that the quantum Zeno effect would result in an exponential slow-down in the general case.\footnote{The method based on the quantum Zeno effect is only efficient if the stationary distributions of adjacent Markov chains are very close to each other.   In contrast, amplitude amplification works as long as the distances of adjacent stationary distributions are bounded from below by the inverse of some polynomial in the input length.  The latter condition is referred to as slowly-varying.}  This is avoided by amplitude amplification in our approach, which can lead to an overall speed-up in the general case.

%

The paper is organized as follows. In Section~2 we present a modified version of Szegedy's quantum analog of an ergodic reversible Markov chain whose unique eigenvector with eigenvalue $1$ (on the relevant computational subspace) is the quantum sample of the stationary distribution.  We also describe its spectral properties in detail.  In Section~3 we present a primitive for preparing quantum states based on amplitude amplification and also a primitive for implementing approximately phase gates that are needed for amplitude amplification.  In Section~4 we use the primitives to obtain a quantum method for preparing the quantum sample of an arbitrary reversible Markov chain provided that we have a sequence of slowly-varying Markov chains.  In Section~5 we show that our approach makes it possible to prepare quantum samples of Boltzmann-Gibbs distributions more efficiently.  

\section{Quantum analog of classical ergodic reversible Markov chains}
We refer the reader to \cite{Szegedy,Magniez} for more details on the quantization.  Let $\cH=\C^N \otimes \C^N$.  The basis states of $\cH$ are denoted by $|x y\>$ for $x,y\in\Omega$.  For $x\in\Omega$, define the normalized vectors
\[
|p_x\>=\sum_{y\in\Omega} \sqrt{p_{xy}} |y\>\,.
\]
where $p_{xy}$ denotes the transition probability from $x$ to $y$.  A {\em quantum update} is any unitary $U$ that satisfies
\begin{equation*}
U|x\>|0\> = |x\> |p_x\>
\end{equation*}
for some fixed state $0\in\Omega$ and all $x\in\Omega$. We refer to the cost to realize $U$ and its inverse $U^\dagger$ as the quantum update cost.

To construct the quantum walk, we define the subspaces
\begin{eqnarray*}
\cA         & = & \s\{|x\>|0\> \,:\, x\in\Omega\} \\
\cB         & = & U^\dagger S U \cA\,,
\end{eqnarray*}
where $S$ denotes that the unitary operator swapping the two tensor components of $\cH$. For $\cK=\cA,\cB$, denote by $\Pi_\cK$ the orthogonal projection onto $\cK$ and by 
\[
R_\cK=2\Pi_\cK-I\,.
\]
the reflection around $\cK$.

\begin{definition}[Quantum walk]
The quantum walk $W(P)$ based on the classical reversible Markov chain $P$ is defined to be the unitary operation (rotation)
\begin{equation}\label{eq:WP}
W(P) = R_\cB \cdot R_\cA
\end{equation}
\end{definition}

\medskip
The quantum walk $W(P)$ can be realized by applying both $U$ and $U^\dagger$ twice:
\[
W(P) =  U^\dagger \cdot S \cdot U \cdot R_\cA \cdot U^\dagger \cdot S \cdot U \cdot R_\cA\,.
\]
Our definition of $W(P)$ is equal to that in \cite{Somma}.  This is different from the definition used in \cite{Szegedy,Magniez} 
\begin{equation*}
\tilde{W}(P) = R_{\tilde{\cB}} \cdot R_{\tilde{\cA}}
\end{equation*}
where
\begin{eqnarray*}
\tilde{\cA} & = & \s\{|x\>|p_x\> \,:\, x\in\Omega\} = U \cA \\
\tilde{\cB} & = & \s\{|p_x\>|x\> \,:\, x\in\Omega\} = U \cB\,.
\end{eqnarray*}
Since $W(P)$ and $\tilde{W}(P)$ are equal up to conjugation by $U$, we can apply the spectral analysis from \cite{Szegedy} to determine the spectrum of $W(P)$. We refer to the subspace $\cA + \cB$ as the busy subspace and to its orthogonal complement, i.e., $\cA^\perp\cap\cB^\perp$, as the idle subspace. Clearly, the operator $W(P)$ acts as identity on the idle subspace.  On the busy subspace, the spectrum of $W(P)$ is as follows.

\begin{theorem} Let $P$ be a time-reversible Markov chain.  Let $\theta_1,\ldots,\theta_{M}\in (0,\frac{\pi}{2})$ be such that $|\lambda_1|=\cos(\theta_1),\ldots,|\lambda_M|=\cos\theta_{M}$ where $M\le N-1$ and the remaining eigenvalues are equal to $0$, i.e., $\lambda_{M+1},\ldots,\lambda_{N-1}=0$.
\begin{enumerate}
\item 
On $\cA \cap \cB$ the operator $W(P)$ acts as the identity $I$. This subspace is one dimensional and is spanned by the eigenvector $|\pi\>|0\>$ where
\begin{equation*}
|\pi\> = \sum_{x} \sqrt{\pi_x} |x\>
\end{equation*}
is the quantum sample of the stationary distribution $\pi$ of $P$.
\item 
On $\cA \cap \cB^\perp$ and $\cA^\perp \cap \cB$ the operator $W(P)$ acts as $-I$.  The dimensions of $\cA\cap\cB^\perp$ and $\cA^\perp\cap\cB$ are equal to $N-1-M$, i.e., the dimension of the kernel of $P$.
\item 
On $\cA+\cB$ those eigenvalues of $W(P)$ that have non-zero imaginary part are exactly $e^{\pm 2i\theta_1},\ldots,
e^{\pm 2i\theta_M}$ with the same multiplicity.
\item 
$W(P)$ has no other eigenvalues on $\cA + \cB$.
\end{enumerate}
\end{theorem}
\noindent
{\bf Proof:} This follows from \cite[Theorem~4]{Magniez} and the results in \cite[Section~12]{Szegedy}. 

\medskip
In abuse of notation, we often use $|\pi\>$ instead of $|\pi\>|0\>$.  In the following, we always stay in the busy subspace $\cA+\cB$.  This is important because we want to obtain the unique eigenvector $|\pi\>$ and not any other eigenvector with eigenvalue $1$ contained in the idle subspace.

The phase gap of $\Delta(P)$ of $W(P)$ is defined to be $2\theta_1$. This is motivated by the above theorem since the angular distance of $1$ from any other eigenvalue (corresponding to an eigenvector in the busy subspace) is at least $\Delta(P)$. The phase gap satisfies 
\begin{equation*}
\Delta(P) \ge |1-e^{2i\theta_1}| = 2\sqrt{1-\lambda_1^2} \ge 2\sqrt{\delta(P)}\,.
\end{equation*}
This inequality is at the heart of the quadratic speed-up due to quantum walks.

\section{Primitives}

\subsection{Preparation via amplitude amplification}
We use Grover's $\frac{\pi}{3}$-amplitude amplification (fixed point search), making it possible to drive a source state to the desired target state by applying a sequence of phase gates \cite{fixedPoint}.  It is a special case of the general approach to amplitude amplification based on the phase matrix that was introduced in \cite{Hoyer}.
\begin{lemma}\label{lem:primitiveOne}
Let $|t_i\>$ and $|t_{i+1}\>$ be two arbitrary quantum states in $\C^d$ with $|\<t_i|t_{i+1}\>|^2\ge p$ for some $p$ with $0<p\le 1$. Denote by $\Pi_i$ the projection on the subspace spanned by $|t_i\>$ and by $\Pi_{i+1}^\perp$ the projection onto the orthogonal subspace. Let $\omega=e^{\frac{\pi}{3}i}$. Define the unitaries
\begin{eqnarray*}
R_i     & = & \omega\Pi_i     + \Pi_i^\perp     \\
R_{i+1} & = & \omega\Pi_{i+1} + \Pi_{i+1}^\perp\,.
\end{eqnarray*}
Define the unitaries $U_{i;m}$ recursively as follows:
\begin{eqnarray*}
U_{i;0}   & = & I \\
U_{i;m+1} & = & U_{i;m} \,\cdot\, R_i \,\cdot\, U_{i;m}^\dagger \,\cdot\, R_{i+1} \,\cdot\, U_{i;m} \\
\end{eqnarray*}
Then, at the $m$th level of recursion we have
\begin{equation*}
|\<t_{i+1}|U_{i,m}|t_i\>|^2 \ge 1 - (1-p)^{3^m}\,.
\end{equation*}
The unitaries in $\{R_i,R_i^\dagger,R_{i+1}, R_{i+1}^\dagger\}$ are used at most $3^m$ times.
\end{lemma}

\medskip
\noindent
Note that the running time of the algorithm is worse than that of Grover's algorithm (the latter is also a special case of amplitude amplification based on the phase matrix).  The reason why we cannot use Grover's algorithm is as follows.  Since we only have a lower bound on the overlap between $|t_i\>$ and $|t_{i+1}\>$, we do not know how many Grover iterations we have to apply without overshooting.  Therefore, we have to employ the version of Grover's algorithm considered in \cite{Boyer}, making it possible to obtain the target state even if the overlap is not known.  However, the problem is that for this algorithm we have to prepare the initial state {\em several} times.  This prevents us from using Grover's algorithm because it is absolutely necessary for our primitive discussed below that the initial state is prepared only once.  

\medskip
\begin{cor}\label{cor:perfectChained}
Let $|t_0\>,\ldots,|t_r\>$ be arbitrary quantum states in $\C^d$ with $|\<t_i|t_{i+1}\>|^2\ge p$ for $i=0,\ldots,r-1$.  Given the state $|t_0\>$, we can prepare a state $|\tilde{t}_r\>$ such that 
\begin{equation*}
\big\| |\tilde{t}_r\> - |t_r\> \big\| \le \epsilon_1\,,
\end{equation*}
for any $\epsilon_1>0$, by invoking the unitaries from $\{R_i,R_i^\dagger\,:\,i=0,\ldots,r\}$ no more than 
\begin{equation*}
L = \frac{12r \log(2r/\epsilon_1)}{\log\big(1/(1-p)\big)}
\end{equation*}
times.
\end{cor}
\noindent
{\bf Proof:} Set $q=1-p$ and $M=3^m$. For $i=0,\ldots,r-1$, define $|t'_{i+1}\>=U_{i;m}|t_i\>$. It follows from Lemma~1 that $|t'_{i+1}\>$ can be expressed as
\begin{equation*}
|t'_{i+1}\> = \alpha |t_{i+1}\> + \beta |t_{i+1}^\perp\>
\end{equation*}
where $\alpha$, $\beta$ are two probability amplitudes with $|\alpha|\ge\sqrt{1-q^M}$, $|\beta|\le\sqrt{q^M}$, and $|t_{i+1}^\perp\>$ is some state with $\<t_{i+1}^\perp|t_{i+1}\>=0$.  Consequently, we have 
\begin{equation*}
\big\| |t'_{i+1}\> - |t_{i+1}\> \big\| \le 1 - \sqrt{1-q^M} + \sqrt{q^M} \le 2\sqrt{q^M}\,.
\end{equation*}
For $i=0,\ldots,r-1$, define
\begin{equation*}
|\tilde{t}_{i+1}\> = \prod_{j=0}^{i} U_{j;m} |t_0\>\,.
\end{equation*}
The task is now to show how to choose $m$ so that
\begin{equation*}
\big\| |t_r\> - |\tilde{t}_r\> \big\|  \le  \epsilon_1\,.\\
\end{equation*}
To do this, we use induction.  The base step is
\begin{equation*}
\big\| |\tilde{t}_1\> - |t_1\> \big\| \le 2\sqrt{q^M}\,.
\end{equation*}
The inductive step is
\begin{eqnarray*}
&& \big\| |\tilde{t}_r\> - |t_r\> \big\| \\
& = &
\big\| |\tilde{t}_r\> - U_{r-1;m} |t_{r-1}\> + U_{r-1;m} |t_{r-1}\> - |t_r\> \big\| \\
& \le &
\big\| U_{r-1;m} \big\| \, \cdot \, \big\| |\tilde{t}_{r-1}\> - |t_{r-1}\> \big\| + \big\| |t'_r\> - |t_r\> \big\| \\
& \le &
\big\| |\tilde{t}_{r-1}\> - |t_{r-1}\> \big\| + 2 \sqrt{q^M}\,.
\end{eqnarray*}
We obtain
\[
\big\| |\tilde{t}_r\> - |t_r\> \big\| \le 2r \sqrt{q^M}\,.
\]
To make the norm distance less or equal to $\epsilon_1$, it always suffices to choose $M$ to be the smallest power of $3$ satisfying
\begin{equation*}
M \ge \frac{2\log(2r/\epsilon_1)}{\log\big(1/(1-p)\big)} \,.
\end{equation*}
It follows that the unitaries from the set $\{R_i,R_i^\dagger\,:\, i=0,\ldots,r+1\}$ are used at most $2rM$ times.  This number is bounded from above by
\begin{equation*}
L = \frac{12 r\log(2r/\epsilon_1)}{\log\big(1/(1-p)\big)} \,.
\end{equation*}

\subsection{Approximate phase gates}
In this section we consider the case where the states $|t_0\>, |t_1\>, \ldots, |t_r\>$ in Corollary~1 are quantum samples of stationary distributions.  We show how to approximately implement the required phase transformation using quantum walks and a variant of the phase estimation algorithm.
 
\begin{lemma}
Let $W$ be a unitary acting on $\C^d$ with unique eigenvector $|\psi_0\>$ with eigenvalue $\lambda_0=1$. Denote the remaining eigenvectors and eigenvalues of $W$ by $|\psi_j\>$ and $\lambda_j=e^{2\pi i \varphi_j}$ for $j=1,\ldots,d-1$, respectively. Let
\[
\Delta=\min_{j=1,\ldots,d-1} |\varphi_j|
\]
be the phase gap of $W$. Let 
\begin{eqnarray*}
a & = & \big\lceil \log (1/\Delta)\big\rceil \\
c & = & \big\lceil \log(1/\sqrt{\epsilon_2})\big\rceil
\end{eqnarray*}
for some $\epsilon_2>0$. Then, there is a quantum circuit $V$ acting on $\C^d\otimes{(\C^2)}^{\otimes ac}$ that invokes the controlled-$W$ gate at most $2^a \cdot c$
times and has the following properties
\begin{eqnarray*}
V |\psi_0\>|0\>^{\otimes ac} & = & |\psi_0\> |0\>^{\otimes ac} \\
V |\psi_j\>|0\>^{\otimes ac} & = & \sqrt{1-\epsilon_2} \, |\psi_j\>|\chi_j\> + \sqrt{\epsilon_2} \, |\psi_j\>|0\>^{\otimes ac} 
\end{eqnarray*}
where $|\chi_j\>$ are some unit vectors in ${(\C^2)}^{\otimes ac}$ with $\<0\cdots 0|\chi_j\>=0$ for $j=1,\ldots,d-1$.
\end{lemma}
\noindent
{\bf Proof:} First, we apply the phase estimation circuit $U$ with $a$ ancilla qubits as depicted below. The circuit $U$ invokes the controlled-$W$ gate $2^a-1$ times.

\medskip
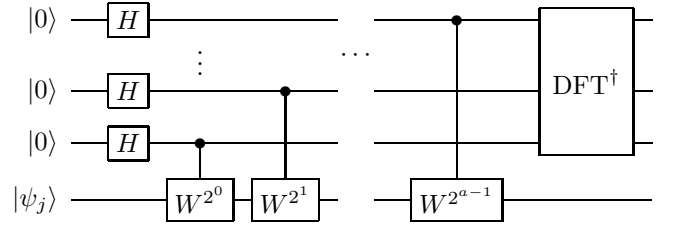
\hspace{0.25cm}
\Qcircuit @C=0.7em @R=0.7em {
\lstick{|0\>}      & \qw    & \gate{H}    & \qw            & \qw            & \qw    &  &     & \qw & \ctrl{4}       & \qw & \multigate{3}{{\rm DFT}^\dagger} & \qw \\
                   &        &             & \vdots         &                &        & \cdots &     &     &                &     &        &  \\
\lstick{|0\>}      & \qw    & \gate{H}    & \qw            & \ctrl{2}       & \qw    &  &     & \qw & \qw            & \qw & \ghost{{\rm DFT}^\dagger}        & \qw \\ 
\lstick{|0\>}      & \qw    & \gate{H}    & \ctrl{1}       & \qw            & \qw    &  &     & \qw & \qw            & \qw & \ghost{{\rm DFT}^\dagger}        & \qw \\
\lstick{|\psi_j\>} & \qw    & \qw         & \gate{W^{2^0}} & \gate{W^{2^1}} & \qw    &  &     & \qw & \gate{W^{2^{a-1}}} & \qw & \qw                      & \qw
}

\vspace{1cm}
\noindent
We have
\begin{eqnarray*}
&&    
U|\psi_j\>|0\cdots0\>^{\otimes a} \\ 
& = & 
|\psi_j\> \otimes {\rm DFT}^\dagger \left( \frac{1}{\sqrt{2^a}} \sum_{m=0}^{2^a-1} e^{2\pi i m \varphi_j} |m\> \right) \\
& = &
|\psi_j\> \otimes \frac{1}{2^a} \sum_{m,m'=0}^{2^a-1} e^{2\pi i m \varphi_j} \, e^{-2\pi i m m' / 2^a} |m'\>\,.
\end{eqnarray*}
The amplitude $\alpha_{m'}$ of the state $|m'\>$ is
\begin{equation*}
\frac{1}{2^a} \sum_{m}^{2^a-1} e^{2\pi i (\varphi_j - m'/2^a) m} =
\frac{1}{2^a} \frac{1-e^{2\pi i (2^a \varphi_j - m')}}{1-e^{2\pi i (\varphi_j - m'/2^a)}}\,.
\end{equation*}
Observe that for $j=0$
\begin{equation*}
\alpha_{m'} = \left\{
\begin{array}{cc}
1 & \mbox{ if $m'=0$}     \\
0 & \mbox{ if $m'\neq 0$}
\end{array}
\right.
\end{equation*}
and so
\begin{equation*}
U|\psi_0\>|0\>^{\otimes a} = |\psi_0\>|0\>^{\otimes a}\,.
\end{equation*}
Now consider the case $j\neq 0$.  To bound $|\alpha_0|$, we use the inequality $|1-e^{ix}|\ge 2|x|/\pi$ whenever $-\pi\ge x\ge \pi$.  We obtain
\begin{eqnarray*}
|\alpha_0| 
& = & 
\frac{1}{2^a} \,\cdot\, \left| \frac{1-e^{2\pi i 2^a \varphi_j}}{1-e^{2\pi i \varphi_j}} \right| \\
& \le &
\frac{1}{2^{a-1}} \,\cdot\, \left| \frac{1}{1-e^{2\pi i\varphi_j}} \right| \\
& \le &
\frac{1}{2^{a-1}} \,\cdot\, \frac{\pi}{2\cdot 2\pi |\varphi_j|} \\
& = &
\frac{1}{2^{a+1} \,|\varphi_j|} \\
& \le &
\frac{1}{2^{a+1}\Delta} \\
& \le &
\frac{1}{2}\,.
\end{eqnarray*}
We conclude that for $j\neq 0$ we have
\begin{equation*}
U |\psi_j\>|0\>^{\otimes a} = \alpha |\psi_j\>|\chi_j\> + \alpha_0 |\psi_j\> |0\>^{\otimes a}\,,
\end{equation*}
where $\<0\cdots 0|\chi_j\>=0$, $|\alpha|>\sqrt{3}/2$, and $|\alpha_0|\le 1/2$.

Since we are only interested in the amplitude of the state $|0\>^{\otimes a}$ on the ancilla qubits, we can replace the inverse discrete Fourier transform ${\rm DFT}^\dagger$ in the phase estimation circuit by the Walsh-Hadamard transform $H^{\otimes a}$.  This is seen as follows.  Observe that both $H^{\otimes a}$ and ${\rm DFT}$ create the uniformly weighted superposition of all computational basis states when applied to $|0\>^{\otimes a}$, implying that 
\begin{equation*}
\<0\cdots 0| {\rm DFT}^\dagger |\phi\> = \<0\cdots 0| H^{\otimes a} |\phi\> 
\end{equation*}
for an arbitrary state $|\phi\>$ on the ancilla register. 

Second, we reduce the ``error amplitude'' to $\sqrt{\epsilon_2}$ by applying the circuit $U$ $c$ times, using a new block of $a$ ancillas each time.  Let $V$ be the resulting circuit.  $V$ invokes the controlled-$W$ gate $(2^a-1) c$ times.  For $j\neq 0$, we have
\begin{equation*}
V |\psi_j\>|0\>^{\otimes ac} = \sqrt{1-\epsilon_2} |\psi_j\>|\chi_j\> + \sqrt{\epsilon_2} |\psi_j\> |0\>^{\otimes ac}
\end{equation*}
as desired. For $j=0$, we have
\begin{equation*}
V |\psi_0\>|0\>^{\otimes ac} = |\psi_0 |0\>^{\otimes ac}\,.
\end{equation*}
This completes the proof.

\medskip
\begin{cor}
Let $W$ be a unitary acting on $\C^d$ with unique eigenvector $|\psi_0\>$ with eigenvalue $\lambda_0=1$. Denote the remaining eigenvectors and eigenvalues of $W$ by $|\psi_j\>$ and $\lambda_j=e^{2\pi i \varphi_j}$ for $j=1,\ldots,d-1$, respectively.  Let
\[
\Delta=\min_{j=1,\ldots,d-1} |\varphi_j|
\]
be the phase gap of $W$.  Let $\Pi$ be the projector onto the space spanned by $|\psi_0\>$ and $\Pi^\perp$ the projector onto the orthogonal complement.  Let $R$ be the unitary that acts on $\C^d$ as follows
\begin{equation*}
R = \omega \Pi + \Pi^\perp\,.
\end{equation*}
Let 
\begin{eqnarray*}
a & = & \big\lceil \log (1/\Delta) \big\rceil \\
c & = & \big\lceil \log(1/\sqrt{\epsilon_2}) \big\rceil
\end{eqnarray*}
for some $\epsilon_2>0$. Then, there is a quantum circuit $\tilde{R}$ acting on $\C^d\otimes (\C^2)^{\otimes ac}$ that invokes the controlled-$W$ gate 
$2^{a+1} \, \cdot\,  c$ times and has the following properties: for $j=0$,
\begin{equation*}
\tilde{R} |\psi_0\> |0\>^{\otimes ac} = 
\big( R |\psi_0\> \big) |0\>^{\otimes ac}
\end{equation*}
and for $j\neq 0$,
\begin{equation*}
\tilde{R} |\psi_j\> |0\>^{\otimes ac}  = \big( R |\psi_j\> \big) |0\>^{\otimes ac} + |\xi\>\,,
\end{equation*}
where $|\xi\>$ is some error vector in $\C^d\otimes (\C^2)^{\otimes ac}$ with 
\begin{equation*}
\| |\xi\> \| \le 2\sqrt{\epsilon_2}\,.
\end{equation*}
\end{cor}
\noindent
{\bf Proof:} Let
\begin{equation*}
\tilde{R} = V^\dagger \, \cdot\, \big( I_d \, \otimes \, Q\big)  \, \cdot\,
V\,,
\end{equation*}
where $Q$ is the following phase gate
\begin{equation*}
Q = \omega |0\>\<0|^{\otimes ac} + \big( I - |0\>\<0|^{\otimes ac} \big)
\end{equation*}
that acts on the ancilla register. 

For $j=0$, it clear that
\[
\tilde{R} |\psi_0\> |0\>^{\otimes ac} = \big( R|\psi_0\>\big) |0\>^{\otimes ac} = \omega |\psi_j\> |0\>^{\otimes ac} \,.
\]
Let us now analyze the action of $\tilde{R}$ for $j\neq 0$.  The state after the application of $V$ is
\[
V |\psi_j\>|0\>^{\otimes ac} = \sqrt{1-\epsilon_2} |\psi_j\>|\chi_j\> + \sqrt{\epsilon_2} |\psi_j\> |0\>^{\otimes ac}
\]
where $\<\chi_j|0\>^{\otimes ac}=0$. The state after the application of $I_d \otimes Q$ is
\begin{eqnarray*}
&& 
\sqrt{1-\epsilon_2} |\psi_j\> |\chi_j\>         + \sqrt{\epsilon_2} \,\omega\, |\psi_j\> |0\>^{\otimes ac} \\
&=&
\sqrt{1-\epsilon_2} |\psi_j\> |\chi_j\>         + \sqrt{\epsilon_2} \,\omega\, |\psi_j\> |0\>^{\otimes ac} + \\
&&
\sqrt{\epsilon_2}   |\psi_j\> |0\>^{\otimes ac} - \sqrt{\epsilon_2}            |\psi_j\> |0\>^{\otimes ac} \\
& = &
|\psi_j\> \otimes \big( \sqrt{1-\epsilon_2} |\chi_j\> + \sqrt{\epsilon_2} |0\>^{\otimes ac} \big) + |\xi'\>\,,
\end{eqnarray*}
where 
\begin{equation*}
|\xi'\> = \sqrt{\epsilon_2} \big(\omega -1 \big) |\psi_j\> |0\>^{\otimes ac}\,.
\end{equation*}
We have $\| |\xi'\> \| \le 2 \sqrt{\epsilon_2}$. In the final step, the application of $V^\dagger$ leads to the state
\begin{equation*}
|\psi_j\>|0\>^{\otimes ac} + V^\dagger |\xi'\> =
|\psi_j\>|0\>^{\otimes ac} + |\xi\>
\end{equation*}
with $\| |\xi\> \| \le 2\sqrt{\epsilon_2}$. We conclude
\begin{equation*}
\tilde{R}|\psi_j\>|0\>^{\otimes ac} = |\psi_j\>|0\>^{\otimes ac} + |\xi\> =
\big( R |\psi_j\>\big) |0\>^{\otimes ac} + |\xi\>\,.
\end{equation*}
It follows from Lemma~2 that $V$ and $V^\dagger$ invoke the controlled-$W$ gate $(2^a-1)c$ times. So, $\tilde{R}$ invokes the controlled-$W$ gate $2(2^a-1)c<2^{a+1}\cdot c$ times.

\section{Quantum sampling}
We now use Corollary~1 and Corollary~2 to prove the following theorem.
\begin{theorem}
Let $P_0,P_1,\ldots,P_r$ be classical Markov chains with stationary distributions $\pi_0,\pi_1,\ldots,\pi_r$ and spectral gaps $\delta_0,\delta_1,\ldots,\delta_r$, respectively. Assume the stationary distributions of adjacent Markov chains are close to each other in the sense that their stationary distributions $\pi_i$ and $\pi_{i+1}$ are close with respect to fidelity, i.e.,
\begin{equation*}
\Big(
\sum_{x\in\Omega} \sqrt{\pi_i(x)} \sqrt{\pi_{i+1}(x)} 
\Big)^2
= 
|\<\pi_i|\pi_{i+1}\>|^2 \ge p
\end{equation*}  
for $i=0,\ldots,r-1$, 
\begin{equation*}
\min \{ \delta_i \,:\, i=0,\ldots,r\} \ge \delta\,,
\end{equation*}
and we can prepare the quantum sample $|\pi_0\>$.  

Then, for any $\epsilon > 0$, there is a quantum sampling algorithm, making it possible to sample according to a probability distribution $\tilde{\pi}_r$ that is close to $\pi_r$ with respect to the total variation distance, i.e., $D(\tilde{\pi}_r,\pi_r) \le \epsilon$.

The algorithm invokes the controlled-$W_i$ operators at most $2^{a+1} \cdot c \cdot L$ times where
\begin{eqnarray*}
L & = & \frac{12 r \log\big(8 r / \epsilon)}{\log\big(1/(1-p)\big)} \\
a & = & \Big\lceil \log(1/\Delta) \Big\rceil \\
c & = & \Big\lceil \log \Big( \frac{96r\log(8r/\epsilon)}{\epsilon \log\big(1/(1-p)\big)} \Big) \Big\rceil\,.
\end{eqnarray*}
\end{theorem}
\noindent
{\bf Proof:} Corollary~1 shows that, given the initial state $|\pi_0\>$, we can prepare a state $|\tilde{\pi}_r\>$ with $\| |\pi_r\> - |\tilde{\pi}_r\> \| \le \epsilon_1$ by invoking the unitaries from the set $\{R_i,R_i^\dagger\,:\,i=0,\ldots,r-1\}$ no more than 
\begin{equation*}
L = \frac{12 r\log(2r/\epsilon_1)}{\log\big(1/(1-p)\big)}
\end{equation*}
times. 

In Corollary~1 we assumed that we can implement these exactly.  However, in reality we can only implement the operators $\tilde{R}_i$ and their inverses $\tilde{R}_i^\dagger$ as described in Corollary~2. This approximation adds an error vector $|\xi\>$ every time an operator $\tilde{R}_i$ or $\tilde{R}_i^\dagger$ is applied, where $\| |\xi\> \|\le 2\sqrt{\epsilon_2}$ for some $\epsilon_2>0$.  

Let $|\tilde{\psi}\>$ be the state obtained by implementing Corollary~2 using $\tilde{R}_i$ to approximate $R_i$. Then, since these operators or their inverses are invoked no more than $L$ times we have
\begin{equation*}
|\tilde{\psi}\> = |\tilde{\pi}_r\>_{\cH}|0\>^{\otimes ac'}_{\cA} + |\xi\>
\end{equation*}
where $c'=\lceil \log(1/\sqrt{\epsilon_2}) \rceil$ and $|\xi\>$ is some vector with 
\begin{equation*}
\| |\xi\> \| \le 2 L \sqrt{\epsilon_2}\,.
\end{equation*}
$\cH$ is the Hilbert space that our quantum samples live in and $\cA$ is the Hilbert space of the ancilla qubits that are required to implement the approximate phase gates $\tilde{R}_i$ and their inverses. 

Let 
\begin{equation*}
|\psi\> = |\pi_r\>|0\>^{\otimes ac}
\end{equation*}
be the ideal state.  We choose $\epsilon_1=\epsilon/4$ and $\epsilon_2=\epsilon^2/(64 L^2)$ so that
\begin{eqnarray*}
\big\| |\psi\> - |\tilde{\psi}\> \big\| 
& \le &
\big\| |\psi\> - |\tilde{\pi}_r\>|0\>^{\otimes ac} \big\| + \big\| |\tilde{\pi}_r\>|0\>^{\otimes ac} - |\tilde{\psi}\> \big\| \\
& \le &
\epsilon_1 + 2L\sqrt{\epsilon_2} \\
& = & \epsilon/4 + \epsilon/4 \\
& = &
\epsilon/2\,.
\end{eqnarray*}
For each $x\in\Omega$, we define the projector 
\begin{equation*}
\Lambda_x = |x\>\<x| \otimes |0\>\<0|^{\otimes ac}
\end{equation*}
acting on $\cH\otimes\cA$.  Let
\begin{equation*}
\Lambda_0 = I_{\cH\cA} - I_{\cH} \otimes |0\>\<0|^{\otimes ac}\,.
\end{equation*}

Let $\Omega'=\Omega\cup\{0\}$.  Observe that the desired distribution $\pi$ is equal to the probability distribution given by
\begin{equation*}
\pi(x)         = \| \Lambda_x  |\psi\> \|^2\,.
\end{equation*}
Our protocol yields the probability distribution 
\begin{equation*}
\tilde{\pi}(x) = \| \Lambda_x  |\tilde{\psi}\> \|^2 \,.
\end{equation*}

We now bound the total variation distance between $\pi$ and $\tilde{\pi}$ from above.  For a subset $S\subseteq \Omega'$, let
\begin{equation*}
\Lambda_S = \sum_{x\in S} \Lambda_x\,.
\end{equation*}
We have
\begin{eqnarray*}
D(\pi,\tilde{\pi})
& = & 
\max_{S\subseteq \Omega'} \big| \pi(S) - \pi(\tilde{S}) \big| \\
& = & 
\max_{S\subseteq \Omega'} \Big| \big\| \Lambda_S |\psi\> \big\|^2 - \big\| \Lambda_S |\tilde{\psi}\> \big\|^2 \Big| \\
& \le &
2  \max_{S\subseteq \Omega'} \Big| \big\| \Lambda_S |\psi\> \big\| - \big\| \Lambda_S |\tilde{\psi}\> \big\| \Big| \\
& \le &
2\, \big\| |\psi\> - |\tilde{\psi}\> \big\| \\
& \le &
\epsilon\,.
\end{eqnarray*}

It follows from Corollaries~1 and 2 that we invoke the controlled-$W_i$ operators or their inverses at most $2^{a+1}\cdot c \cdot L$ times.

\section{Quantum Simulated Annealing}

The Metropolis algorithm refers to a general construction that transforms any irreducible Markov chain on state space $\Omega$ to a time-reversible Markov chain with a required stationary distribution.  We consider the case where the desired stationary distribution is equal to the Boltzmann-Gibbs distribution $\pi_\beta$ of some (classical) Hamiltonian $H$ at (inverse) temperature $\beta$. We show how to prepare the corresponding quantum sample $|\pi_\beta\>$ using Theorem~2.  The resulting algorithm has a better running time than that of the recently proposed algorithm based on the quantum Zeno effect \cite{Somma,Somma2}.

For completeness, we give a short description of the Metropolis algorithm.  The presentation is based on \cite[Lemma 10.8]{Mitzenmacher}. For a finite state space $\Omega$ and neighborhood structure $\{N(x) \, : \, x\in\Omega\}$, let $N=\max_{x\in\Omega} |N(x)|$. Let $M$ be any number such that $M\ge N$.  For all $x\in\Omega$, let $E(x)\ge 0$ be the energy of the state $x$.  The desired stationary distribution is $\pi_\beta = \big( \pi_\beta(x) \, : x\in\Omega \big)$, where 
\begin{equation*}
\pi_\beta(x) = \frac{\exp(-E(x)\beta}{Z_\beta}
\end{equation*}
is the probability of state $x$ and
\begin{equation*}
Z_\beta = \sum_{x\in\Omega} \exp(-\beta E(x))\,,
\end{equation*}
denotes the partition function at temperature $\beta$.  

Consider the Markov chain $P_\beta$ whose transition probabilities $p_{xy}$ are:
\[
\begin{array}{ll}
\min\Big\{1,\exp\big((E(y)-E(x)\big)\beta)\Big\}/M & \mbox{if $x\neq y, y\in N(x)$} \\
0                                                  & \mbox{if $x\neq y, y\not\in N(x)$} \\
1-\sum_{z\neq x} p_{xz}                            & \mbox{if $x=y$}
\end{array}
\]
Then, if this chain is irreducible and aperiodic, $P_\beta$ is time-reversible and its stationary distribution is given by the Boltzmann-Gibbs probabilities $\pi_\beta(x)$. Let $\delta_\beta$ denote its spectral gap.

Let $H$ be the Hamiltonian defined by
\begin{equation*}
H = \sum_{x\in\Omega} E(x) |x\>\<x|\,.
\end{equation*}
We need the following simple lemma that characterizes how the quantum sample changes when the temperature is increased.

\begin{lemma}
The quantum samples of the stationary distributions of the above Metropolis process at temperatures $\beta$ and $\beta+\Delta\beta$ satisfy
\[
|\<\pi_{\beta}|\pi_{\beta+\Delta\beta}\>|^2\ge \exp(-\|H\| \Delta\beta)
\]
for all $\beta$ and all $\Delta\beta$.
\end{lemma}
\noindent {\bf Proof: } We have
\begin{eqnarray*}
&&
\<\pi_{\beta}|\pi_{\beta+\Delta\beta}\> \\
& = &
\sum_{x\in\Omega} 
\frac{\exp({-\beta E(x)/2)}}{\sqrt{Z_\beta}} 
\frac{\exp({-(\beta+\Delta\beta) E(x)/2)}}{\sqrt{Z_{\beta+\Delta\beta}}} \\
& \ge &
\sum_{x\in\Omega} \frac{\exp({-\beta E(x))}}{Z_\beta}
\exp({-E(x) \Delta\beta/2}) \\
& \ge &
\exp({-\|H\| \Delta\beta/2})\,.
\end{eqnarray*}
The first inequality follows from the fact that $Z_\beta>Z_{\beta+\Delta\beta}$. By taking the square, we obtain the desired result. 

\medskip
\noindent
The following corollary follows directly from Theorem~2 by observing that $p=1/e$ if we set $\Delta\beta=1/\|H\|$ in Lemma~3.
\begin{cor}
Let $r = \beta\|H\|$, $\beta_i = i / \|H\|$ for $i=0,\ldots,r$,
\begin{equation*}
\delta \le \min\{ \delta_{\beta_i} \,:\, i=1,\ldots,r\}
\end{equation*}
be a lower bound on the smallest spectral gap and $\Delta$ the phase gap corresponding $\delta$.  Then, for any $\epsilon>0$, there is a quantum algorithm that outputs the states $x$ according to a probability distribution $\tilde{\pi}_{\beta_r}$ with
\begin{equation*}
D\big(\tilde{\pi}_{\beta_r},\pi_{\beta_r}\big) \le \epsilon\,.
\end{equation*}
The algorithm invokes the operators from $\{W_{\beta_i}\}$ at most
\begin{equation*}
2^{a+1} \cdot c \cdot L
\end{equation*}
times where
\begin{eqnarray*}
L & = & \frac{12\beta \|H\| \log\big( 8\beta\|H\| / \epsilon \big)}{\log\big(e/(e-1)\big)}\\
a & = & \big\lceil \log(1/\Delta) \big\rceil \\
c & = & \Big\lceil
\log\Big(
\frac{96\beta\|H\| \log(8\beta\|H\|/\epsilon)}{\epsilon \log(e/(e-1))}
\Big)
\Big\rceil\,.
\end{eqnarray*}
\end{cor}

\begin{lemma}
Let $H$ be a Hamiltonian acting on a state space of cardinality $d$ with spectral gap $\gamma$. Let $\Pi$ be the projector onto the eigensubspace corresponding to the minimal eigenvalue.  Then 
\begin{equation*}
|\<\pi_\beta|\Pi|\pi_\beta\>|^2 \ge 1 - \epsilon_3\,.
\end{equation*}
provided that the inverse temperature satisfies
\begin{equation}\label{eq:boundBeta}
\beta \ge \frac{1}{\gamma} \log\Big( \frac{(1-\epsilon_3)\,d}{\epsilon_3} \Big)\,.
\end{equation}
\end{lemma}
\noindent
{\bf Proof:} The worst case occurs when the ground state $|g\>$ is unique and all other states have energy $\gamma+E(g)$ where $E(g)$ is the ground state energy.  In this case the probability of obtaining the ground state when measuring $|\pi_\beta\>$ in the computational basis is
\begin{equation*}
p = \frac{1}{(d-1) e^{-\gamma\beta}+1}\,.
\end{equation*}
To make this probability greater or equal to $1-\epsilon_3$, it suffices to choose the inverse temperature $\beta$ as in eq.~(\ref{eq:boundBeta}).

\medskip
\noindent
By choosing $\epsilon=1/4$ in Corollary~3 and $\epsilon_3=1/4$ in Lemma~4 we obtain the following corollary.
\begin{cor}
There is a quantum algorithm that outputs a ground state of $H$ with probability greater than $1/2$.  It invokes the operators from $\{W_{\beta_i}\, : \, i=1,\ldots, r\}$ at most 
\begin{equation}\label{eq:run}
\frac{1}{\Delta} \,\cdot\, \frac{\|H\|}{\gamma} \,\cdot\, \log d \,\cdot\, \log\left( \frac{\|H\|}{\gamma} \, \log d \right)
\end{equation}
times.
\end{cor}

\medskip
\noindent
Let us explain how the above algorithm differs from the algorithm based on the quantum Zeno effect \cite{Somma2}.  That algorithm has the running time 
\begin{equation}\label{eq:runSomma}
\frac{1}{\Delta'} \, \cdot \,
\left( \frac{\|H\|}{\gamma} 
\right)^2 
\, \cdot \, 
\log^3 d
\,,
\end{equation}
where $\Delta'$ is the phase gap corresponding to the minimal spectral gap in the sequence of Markov chains.  

Both algorithms make use of Szegedy's quantum walk operators to obtain a speed-up over the classical case due to the quadratic relation between phase gaps and spectral gaps.  The reduction from $(\|H\|/\gamma)^2$ in (\ref{eq:runSomma}) to $\|H\|/\gamma$ in (\ref{eq:run}) is due to the advantage of amplitude amplification over the quantum Zeno effect. Note also that $\Delta'\le\Delta$ because the change in temperature between adjacent Markov chains is $\Delta\beta'=O\big(\gamma/(\|H\|^2 \log d)\big)$ in \cite{Somma2} and $\Delta\beta=1/\|H\|$ in our algorithm.  Roughly speaking, amplitude amplification makes it possible to make ``bigger'' jumps (i.e., bigger changes in temperature) than the quantum Zeno effect, without decreasing the success probability.

{\it Conclusions and Discussions}.--- We have presented a simple quantum algorithm, making it possible to prepare quantum samples of stationary distributions of arbitrary slowly-varying Markov chains. It significantly improves upon a previous algorithm for that purpose based on adiabatic generation \cite{TaShma}.  It also provides a speed-up over a recently proposed method for preparing quantum samples of Boltzmann-Gibbs distributions of classical Hamiltonians \cite{Somma2}.

The authors would like to thank Chen-Fu Chiang for helpful comments.  A.~A. and P.~W. gratefully acknowledge the support by NSF grants CCF-0726771 and CCF-0746600.

\end{document}